%% file: main.tex
\documentclass[conference]{IEEEtran}
\usepackage{cite}

\usepackage{graphicx, xcolor}
\ifCLASSINFOpdf
\else
\fi
\usepackage{amsmath, amsthm}
\usepackage[ruled, linesnumbered, norelsize]{algorithm2e}
\usepackage{amssymb}
\newcommand{\R}{\mathbb{R}}

\usepackage{array}
\usepackage{booktabs, multirow}
\usepackage{float}
\usepackage{hyperref, url}

\hypersetup{
    colorlinks=false,
    pdfborder={0 0 0},
    }

\IEEEoverridecommandlockouts

\begin{document}
\title{H-GCN: A Graph Convolutional Network Accelerator on Versal ACAP Architecture}

\newcommand{\BetterMark}{$^\star$}

\author{
Chengming Zhang\BetterMark\IEEEauthorrefmark{5},
Tong Geng\IEEEauthorrefmark{2}\IEEEauthorrefmark{4}, 
Anqi Guo\IEEEauthorrefmark{3}, 
Jiannan Tian\BetterMark, 
Martin Herbordt\IEEEauthorrefmark{3}, 
Ang Li\IEEEauthorrefmark{2},
Dingwen Tao\BetterMark\IEEEauthorrefmark{5}\thanks{Corresponding author: Dingwen Tao (\url{dingwen.tao@wsu.edu}).}\\
\BetterMark%
School of Electrical Engineering and Computer Science, Washington State University, Pullman, WA, USA\\
\IEEEauthorrefmark{2}%
Mathematics and Computer Science Division, Pacific Northwest National Laboratory, Richland, WA, USA\\
\IEEEauthorrefmark{3}%
Department of Electrical and Computer Engineering, Boston University, Boston, MA, USA\\
\IEEEauthorrefmark{4}%
Department of Electrical and Computer Engineering, University of Rochester, Rochester, NY, USA\\
\IEEEauthorrefmark{5}%
School of Informatics, Computing, and Engineering, Indiana University, Bloomington, IN, USA
}


%

\maketitle

\thispagestyle{plain}
\pagestyle{plain}

\begin{abstract}
Graph Neural Networks (GNNs) have drawn tremendous attention due to their unique capability to extend Machine Learning (ML) approaches to applications broadly-defined as having unstructured data, especially graphs. 
Compared with other Machine Learning (ML) modalities, the acceleration of Graph Neural Networks (GNNs) is more challenging due to the irregularity and heterogeneity derived from graph typologies. Existing efforts, however, have focused mainly on handling graphs' irregularity and have not studied their heterogeneity. 
To this end we propose H-GCN, a PL (Programmable Logic) and AIE (AI Engine) based hybrid accelerator that leverages the emerging heterogeneity of Xilinx Versal Adaptive Compute Acceleration Platforms (ACAPs) to achieve high-performance GNN inference. In particular, H-GCN partitions each graph into three subgraphs based on its inherent heterogeneity, and processes them using PL and AIE, respectively. To further improve performance, we explore the sparsity support of AIE and develop an efficient density-aware method to automatically map tiles of sparse matrix-matrix multiplication (SpMM) onto the systolic tensor array. Compared with state-of-the-art GCN accelerators, H-GCN achieves, on average, speedups of 1.1$\sim$2.3$\times$.
\end{abstract}


%
\IEEEpeerreviewmaketitle

\setlength{\textfloatsep}{6pt}
\setlength\abovecaptionskip{3pt}
\setlength{\abovedisplayskip}{2pt}
\setlength{\belowdisplayskip}{2pt}
\setlength{\abovedisplayshortskip}{2pt}
\setlength{\belowdisplayshortskip}{2pt}

\input{tex/01_Introduction}
\input{tex/02_Background}
\input{tex/03_Related_work}
\input{tex/04_Design}
\input{tex/05_Evaluation}
\input{tex/06_Conclusion}

\section*{Acknowledgment}
This work was partially supported by the National Science Foundation through awards OAC-2034169 and CCF-1919130. This work was also partially supported by the Compute-Flow-Architecture (CFA) project under PNNL’s Data-Model-Convergence (DMC) LDRD Initiative. The Pacific Northwest National Laboratory is operated by Battelle for the U.S. Department of Energy under Contract DE-AC05-76RL01830.



%

\clearpage
\bibliographystyle{IEEEtran}
\bibliography{refs}

\end{document}

%% file: tex/01_Introduction.tex
\section{Introduction}
\label{sec:introduction}

In the past few years, GNNs have achieved great success in many applications such as node classification \cite{zhou2019meta}, link prediction \cite{zhang2018link}, graph classification \cite{errica2019fair}, and clustering \cite{zhang2019attributed}. 
Among various kinds of GNNs, graph convolutional network (GCN) \cite{kipf2016semi, hamilton2017inductive} is one category of models that re-define the notion of convolution for graph data and has attracted substantial efforts from both the industrial and academic communities due to their unique ability to extract latent information from graph data. 
GCNs have various applications, including citation networks \cite{kipf2016semi}, social network analysis \cite{levie2018cayleynets}, chemistry \cite{li2019deepchemstable}, computer vision \cite{monti2017geometric}, and natural language processing \cite{yao2019graph}.

Despite the popularity of GCNs, accelerating GCN inference is still challenging: GCNs inherit the irregular computational pattern and processing dataflow of graph analytics, resulting in inefficiency on CPUs and GPUs. This is due especially to three factors: (1) \textit{irregular data access patterns} due to executing on non-Euclidean data, (2) \textit{workload imbalance} due to skewed distribution of graph degrees, and (3) \textit{hybrid computation patterns} due to diverse features of different GCN phases.
In particular, the \textbf{Aggregation} (or message passing) phase performs vector additions where vectors are fetched with irregular strides, while the \textbf{Combination} (or node embedding) phase can be either dense or sparse-dense matrix multiplication. We describe these two phases in detail in Section \ref{sec:background}.

\begin{figure}[t]
    \centering
    \includegraphics[width=\linewidth]{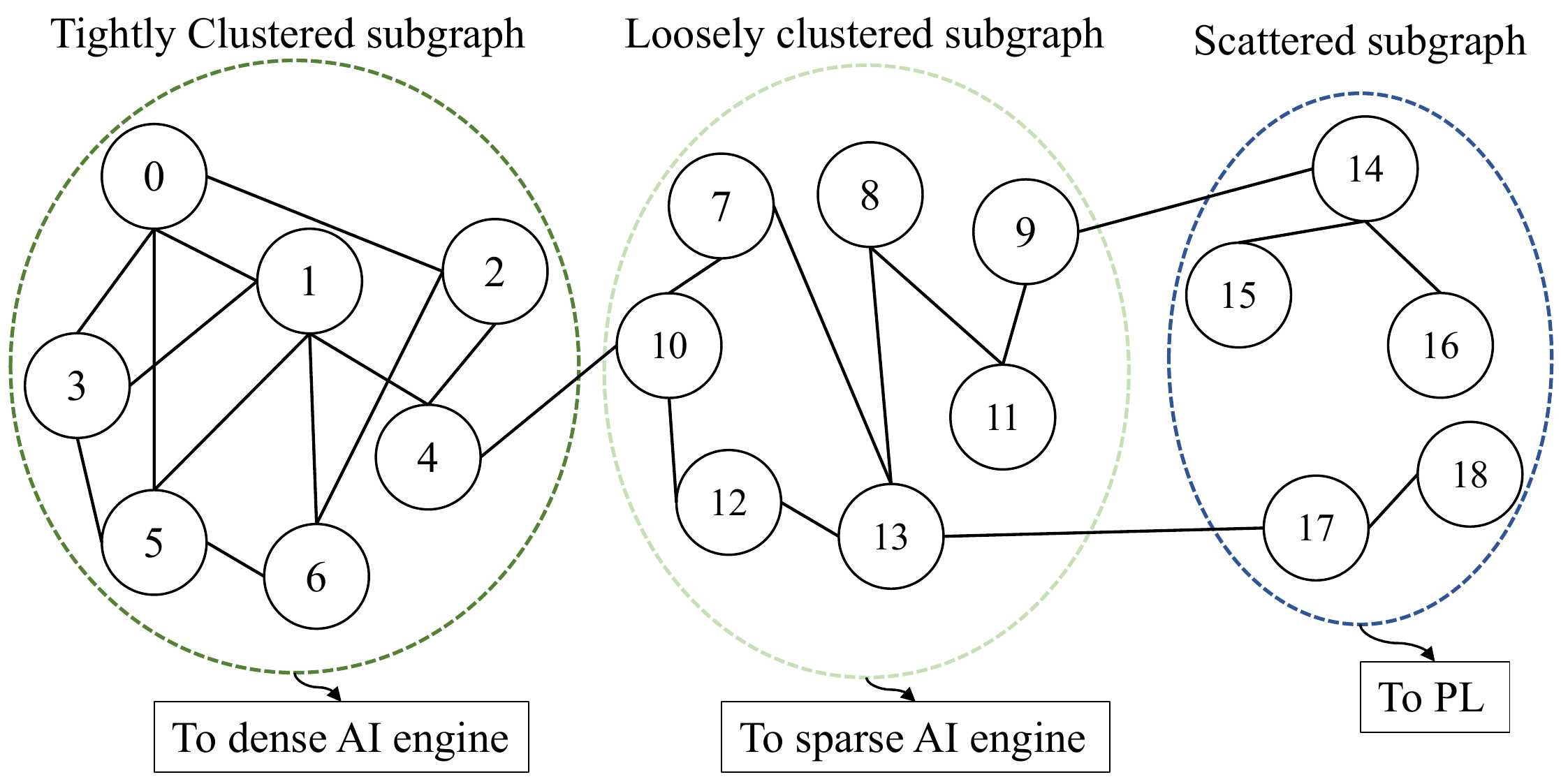}
    \caption{Overview of three types of subgraph.}
    \label{fig:subgraph}
\end{figure}

There have been many efforts on GCN acceleration using both GPUs and FPGAs. Researchers have pointed out that the irregularity from graph topology, the resulting poor data locality, and the serious workload imbalance are the problems \cite{lumsdaine2007challenges}. By leveraging FPGA hardware flexibility, existing work \cite{geng2020awb, zhang2021boostgcn} has well addressed these problems. However, we observe that besides the irregularity, the heterogeneity of graph structure is also a significant performance limiter. As shown in Figure \ref{fig:subgraph}, a graph can have tightly clustered components, loosely clustered components, and scattered nodes: it is therefore challenging to use a unified hardware architecture/device to accelerate all parts of the graph computation.

A few works have implemented GCN accelerator on FPGAs \cite{zeng2020graphact, zhang2021boostgcn}. However, the overall performance is significantly bounded due to the low frequency of FPGAs compared to CPUs and GPUs.
Also, single-instruction multiple-data (SIMD) processing in CPUs can provide high frequency and computation power. Its utility, however, is reduced as the target computation strays from dense, regular operations.
This is also the case to some extent in the analogous modes in GPUs and FPGAs.
Overall, the heterogeneity of GCN implies that emerging heterogeneous hardware such as Xilinx ACAP may provide an opportunity for further acceleration. 

To this end, we propose H-GCN, an accelerator designed to mirror the heterogeneous computing paradigm of GCNs. In particular, H-GCN leverages the heterogeneity of the Versal ACAP to efficiently process different types of subgraphs. The computation of tightly clustered components is mapped onto dense AIEs to fully utilize their high frequency and parallelism from SIMD and very-long instruction word (VLIW) processors. The computation of loosely clustered components is executed on sparse AIEs to reduce computation latency. The computation of scattered nodes is finished on programmable logic (PL) to utilize its programming flexibility. Its performance is not be bounded by the low frequency since the proportion of scattered nodes is relatively small.

In contrast with previous efforts using heterogeneous architectures to process the two GCN phases---Aggregation and Combination---we focus rather on the heterogeneity in the graph itself, which is the fundamental problem in large graph processing.  
To the best of our knowledge, this is the first work that implements a GCN accelerator on real-world heterogeneous hardware ACAPs and tackles sparse tensor computation on the Versal AIEs. Our contributions are summarized as:
\begin{itemize}
  \item We propose H-GCN---an ultra-efficient, systolic tensor-based hardware accelerator---that incorporates the features of the PL and AIE for fully utilizing the ACAP's heterogeneous compute capability in GCN computation.
  \item We study the heterogeneity of graphs and heterogeneity-aware GNN acceleration. 
  \item We are the first to study the use of the AIE compiler in graph processing and sparse matrix processing.
  \item We design a lightweight grouping strategy to enable sparse tensor computation on the Versal AIEs.
  \item We develop an efficient method to process tiles of a sparse matrix to enable an automatic mapping of SpMM onto the systolic tensor array.
  \item Experimental results show that compared with CPU and GPU solutions (i.e., PyG-CPU, PyG-GPU, DGL-CPU, and DGL-GPU), H-GCN achieves up to 3376.3$\times$ and 128.7$\times$ speedups, respectively. Compared with a state-of-the-art FPGA accelerator, H-GCN achieves 1.4$\sim$12.7$\times$ speedup on the tested graph datasets.
\end{itemize}

In Section \ref{sec:background}, we present background about GCN and ACAP. In Section \ref{sec:Related}, we discuss related work on GCN accelerator in detail, comparing them and discussing their limitations.  In Section \ref{sec:design}, we describe our system architecture. In Section \ref{sec:evaluation}, we present the experimental results on various graph datasets. In Section \ref{sec:conclusion}, we conclude and discuss future work.

%% file: tex/02_Background.tex
\section{Background and Motivation}
\label{sec:background}
In this section, we will introduce some background information, including GCNs and Versal ACAPs.

\subsection{Graph Convolutional Networks}
\label{subsec:gcn}
GCNs are composed of stacked graph convolutional layers. Each GCN layer follows the Aggregation and Combination paradigm. 
Particularly, the widely used 2-layer GCN model is
\begin{align}
    X^2 = \sigma \, (\tilde{A} \cdot \sigma \, (\tilde{A} \cdot X^0 \cdot W^1) \cdot W^2),
    \label{equ-1} 
\end{align}
where $W^l \in \R ^{h^{l-1} \times h^l}$ is the weight matrix of the $l^{th}$ layer and $X^l$ is the feature vector of the $l^{th}$ layer. $\tilde{A} = D^{-\frac{1}{2}} \cdot \overline{A} \cdot D^{-\frac{1}{2}}$. Here $\overline{A} = A + I$ is the self-loop adjacency matrix; $D$ is the Laplacian matrix with $D_{ii} = \sum_j \overline{A}_{ij}$; and $\sigma$ denotes non-linear activation functions.

As introduced above, the key computation pattern in GCNs is abstracted into a matrix chain multiplication $A \cdot X \cdot W$. There can be two alternative computation orders: Aggregation first $(A \cdot X) \cdot W$, or Combination first $A \cdot (X \cdot W)$. 
Note that previous works \cite{geng2020awb, liang2020engn} have shown that $A$ is ultra-large and sparse, $X$ is moderate sparse, and $W$ is generally small and dense, thus the ``Combination-first'' approach can better utilize the sparsity of matrix $A$ to reduce arithmetic computation. Consequently, our work H-GCN adopts this Combination-first approach. 

\subsection{Xilinx Versal ACAP}
\label{subsec:acap}

\begin{figure}[t]
    \centering
    \includegraphics[width=1.0\linewidth]{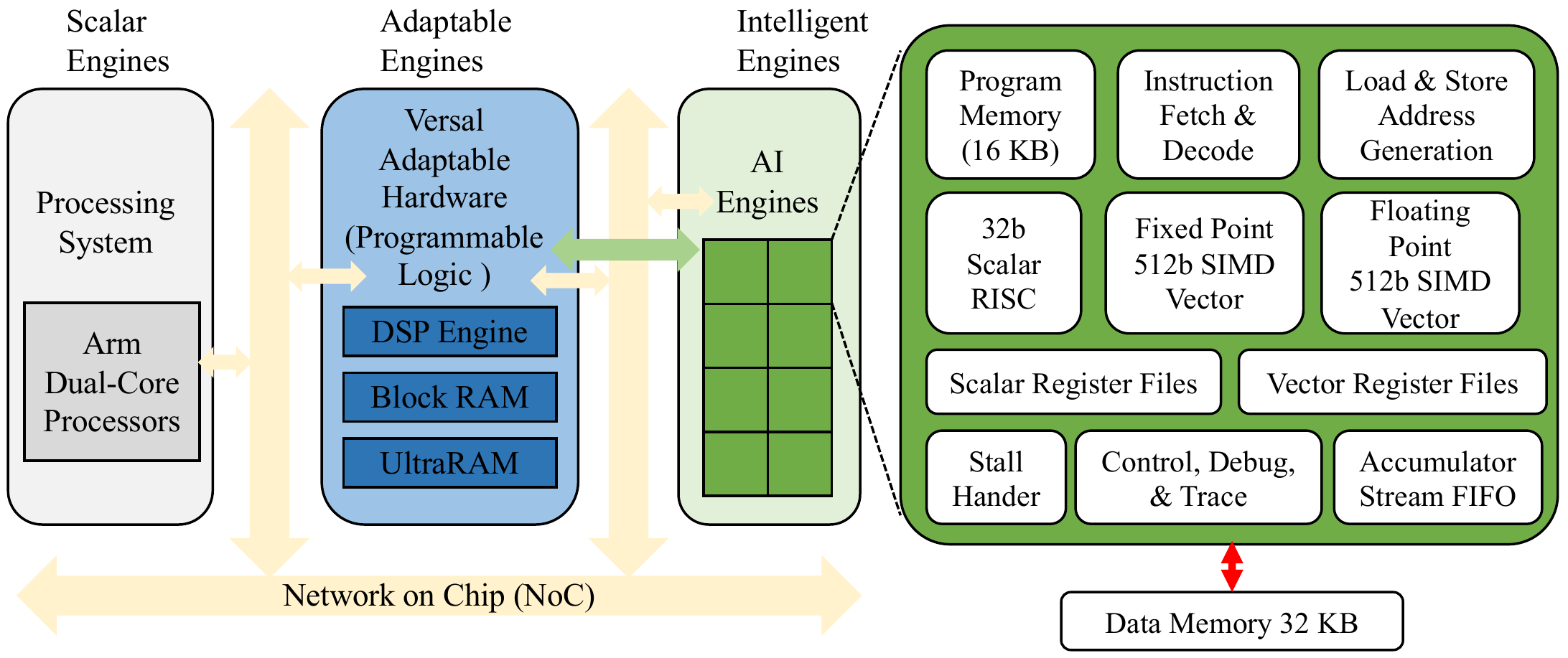}
    \caption{Xilinx Versal Adaptive Compute Acceleration Platforms (ACAPs).}
    \label{fig:acap}
\end{figure}

Figure \ref{fig:acap} shows the Xilinx Versal ACAP architecture. ACAP \cite{gaide2019xilinx, acap} is a fully software-programmable, heterogeneous compute platform that combines three components: (1) the Processor System (PS)---Scalar Engines that include the ARM processors, (2) Programmable Logic (PL)---Adaptable Engines that include the programmable logic blocks and memory, and (3) Artificial Intelligence Engines (AIEs) with leading-edge memory and interfacing technologies.

The PL kernels can be C/C++ kernels or RTL kernels. Its programming model is the same as traditional FPGA. 
Xilinx AIEs are an array of VLIW processors with SIMD vector units, which are highly optimized for compute-intensive applications. 
The AIE array provides three levels of parallelism: (1) SIMD - vector registers that allow multiple elements to be computed in parallel, (2) instruction level - VLIW architecture that allows multiple instructions to be executed in a single clock cycle, and (3) multi-core - AIE array where up to 400 AIEs can execute in parallel.
The AIE kernels are C/C++ programs written using specialized intrinsic calls \cite{aieint} or AIE APIs \cite{aieapi} for the VLIW processor. 
In this work, we mainly use intrinsic calls to implement our AIE kernels and use the Vitis AI compiler ``AIE'' to compile these codes.

In general, if we compare ACAP to a conventional computing system, the PS plays the role of CPU, the PL implements all the FPGA functions, and the AIEs are responsible for the computational acceleration like GPU. Thus, ACAP illustrates a strong heterogeneity. However, there is no work that takes advantage of such strong heterogeneity in GCN acceleration. In addition, intrinsic calls or APIs are designed and optimized for dense computation, so there is no prior work that optimizes sparse computation on the AIEs. 

%% file: tex/03_Related_work.tex
\section{Related Work}
\label{sec:Related}

There have been ongoing researches on designing dedicated hardware architecture to accelerate GCNs.
For example, HyGCN \cite{yan2020hygcn} designs hybrid architecture with individual modules for Aggregation and Combination, respectively, to tackle the hybrid computing pattern of Graph Neural Networks.
AWB-GCN \cite{geng2020awb} proposes an autotuning strategy to solve the workload imbalance in GCN acceleration.
BoostGCN \cite{zhang2021boostgcn} uses hardware-aware partition centric feature aggregation scheme to increase on-chip data reuse. 
I-GCN \cite{geng2021gcn} reorders graphs using islandization to improve the data locality so as to achieve better on-chip data reuse and less off-chip memory access. Islandization targets low frequency, fine-grained, high flexible PL devices and requires fine-grained hardware architecture, which is not suitable for 2D-mesh AIEs.
In the evaluation, we will compare our work with HyGCN, AWB-GCN, BoostGCN, and I-GCN.

Different from all prior work, our proposed H-GCN can fully enable the computational power of the emerging heterogeneous compute platform---Xilinx Versal ACAP---for GCN acceleration by leveraging its strong heterogeneity (e.g., ARM processor, FPGA, and SIMD vector units).
To fully explore the capability of ACAP, we propose to mix sparse/dense systolic tensor arrays to accelerate the hybrid computing pattern of GCNs. We will describe our detailed design in Section \ref{sec:design}. 

In addition, there are a few applications that already leveraged Versal ACAPs.
For example, Corradi and Jensen \cite{corradi2020real} implemented real-time synthetic aperture and plane wave ultrasound imaging on the AIEs.
However, there has been no work that explores the way to implement and optimize sparse computation on AIEs.

%% file: tex/04_Design.tex
\section{System Architecture}
\label{sec:design}
In this section, we introduce the architecture of our proposed H-GCN, followed by the architecture of the sparse tensor engine for feature aggregation. We then explain the design of systolic tensor array for feature update in detail.

\subsection{Overview of Our Proposed Architecture}

\begin{figure}[!t]
    \centering
    \includegraphics[width=\linewidth]{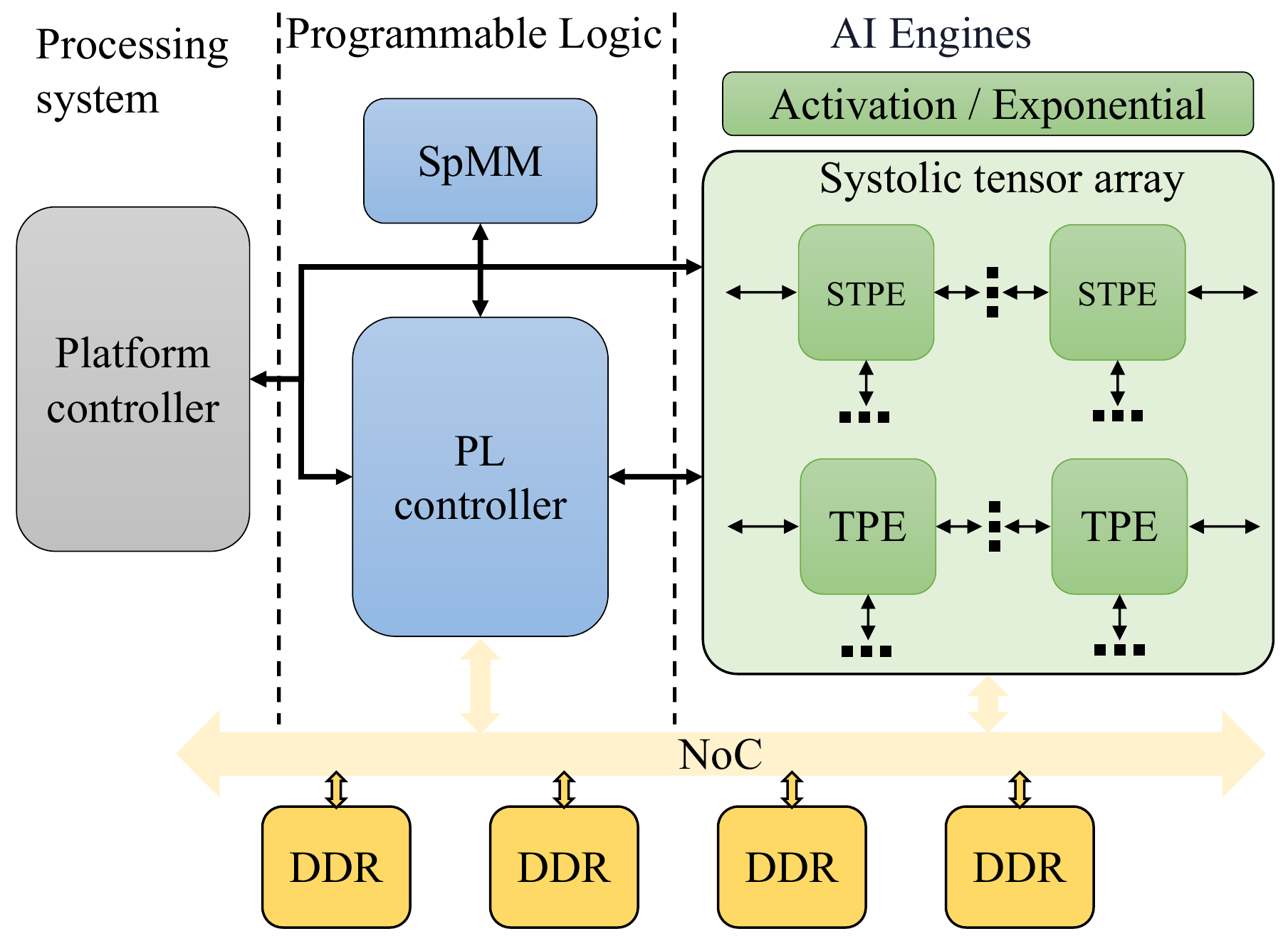}
    \caption{Overview of our hardware system design.}
    \label{fig:arch0}
\end{figure}

Figure \ref{fig:arch0} shows the overview architecture of our proposed H-GCN. It consists of a platform controller in processing system, a sparse-dense matrix-matrix multiplications (SpMM) unit and a PL controller in programmable logic, a sparse/dense systolic tensor array and activation/exponential unit implemented in the AIEs, an network on chip (NoC), four DDR4 SDRAM.
The platform controller is used to control the whole system, send instructions to the SpMM unit, PL controller, and sparse/dense systolic tensor array to control their executions, and collect their statuses.
Specifically, the PL controller controls SpMM unit to cooperate with the sparse/dense systolic tensor arrays to perform all GCN computations. It starts the SpMM unit when it detects that the sparse or dense systolic tensor array has generated enough data. We were inspired by MatRaptor \cite{srivastava2020matraptor} to design our SpMM unit, which adopts row-wise product approach. The PL controller also includes a DDR controller to work with the NoC to perform data reading and writing.
Moreover, the sparse/dense systolic tensor array, which is interconnected side-by-side in a chain/ring fashion, targets the acceleration of both dense and sparse matrix addition and multiplication. It includes both sparse systolic tensor array and dense systolic tensor array; the sparse systolic tensor array is designed for sparse-dense matrix-matrix multiplications in GCNs, while the dense systolic tensor array is mainly for dense-dense matrix-matrix multiplications in GCNs.
In addition, our system first performs graph reordering (Section \ref{subsec:reordering}) to improve the data locality/reuse and then maps different computations, i.e., dense matrix-matrix multiplication and our optimized SpMM (Section \ref{subsec:sparse-tensor} and \ref{subsec:sparse-systolic}), onto different computation engines, i.e., AIEs and PL, based on the matrix density (will be detailed in Section \ref{subsec:setup}).

\subsection{Input Graph Reordering}
\label{subsec:reordering}
Graph reordering is to optimize both the computation order and the data layouts (e.g., graph-level data locality \cite{arai2016rabbit}) by modifying the order of vertices. Our goal of reordering is to group the vertices with more shared neighbors together to improve the data reuse when conducting aggregation reductions. 
The intrinsic reason that the reordering method can provide better temporal reuse is based on the fact that real-world graphs exhibit a ``community'' structure \cite{girvan2002community}, which means some vertices may share neighbors or have a closer relationship to each other; thus, by grouping them together, the data locality during execution will be significantly improved. 
Note that graph reordering does not change the graph structure but only affects the execution order in the graph.

In this work, we perform the graph reordering and sort vertices into a community based on their degrees \cite{chiang2019cluster} at the training stage for only once using mt-metis \cite{lasalle2013multi}. mt-metis is the latest release of an OpenMP version of Metis partitioning and ordering routines. 
More discussion about this overhead will be in Section \ref{sec:evaluation}. 

\begin{figure}[!t]
    \centering
    \includegraphics[width=\linewidth]{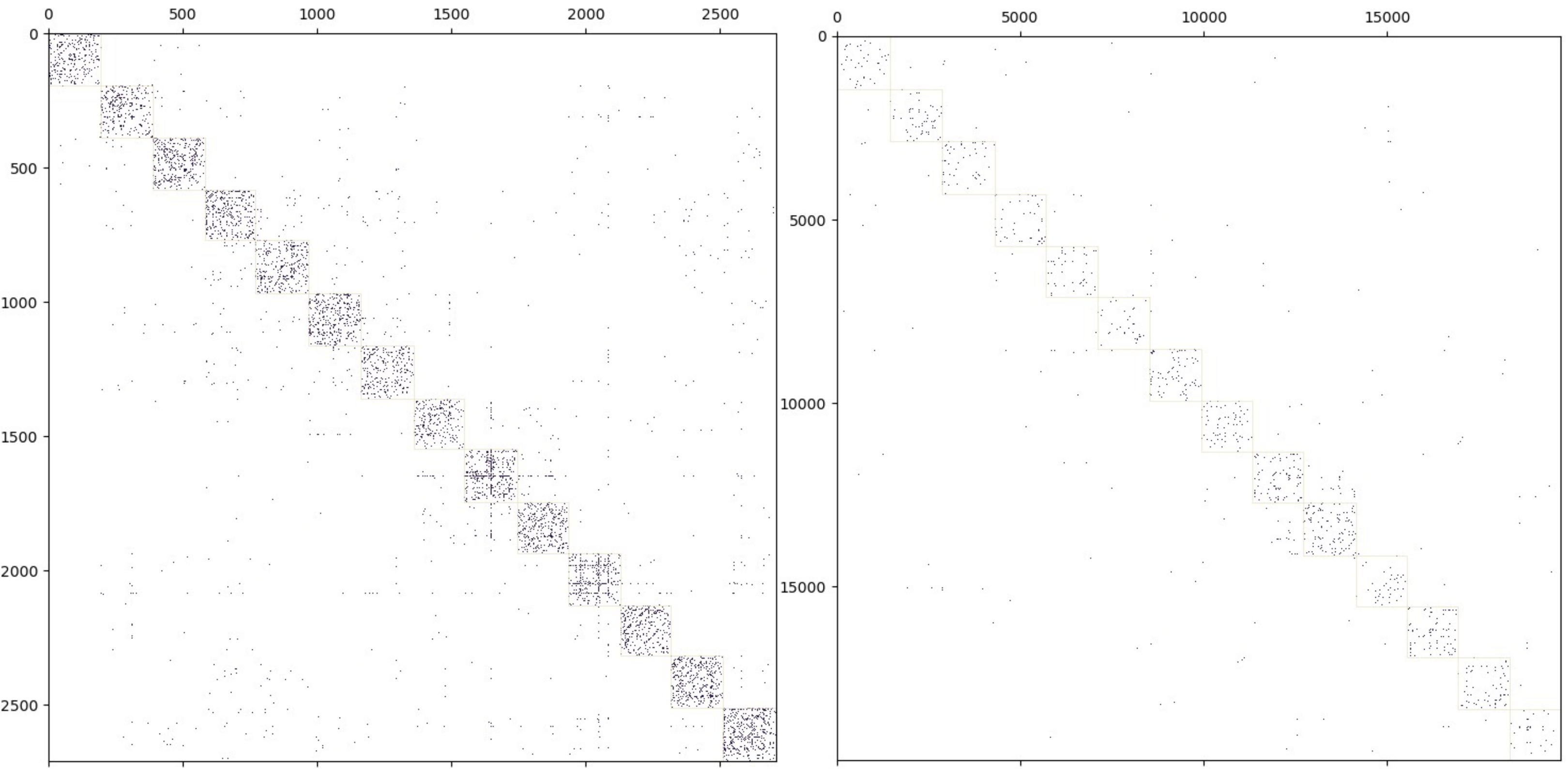}
    \caption{The effect of reordering on Cora (left) and Pubmed (right) \cite{bojchevski2017deep}.}
    \label{fig:reorder}
\end{figure}

Figure \ref{fig:reorder} shows the effect of reordering on the Cora and Pubmed dataset \cite{bojchevski2017deep}. It illustrates that most of the vertices are concentrated in the diagonal area forming relatively dense rectangular areas (each dense area is marked with an auxiliary line in the figure). The effect of concentrating vertices in rectangular areas has three advantages: (1) The potential of data reuse is increased. (2) The denser the data distribution, the higher the computational efficiency of the AIEs. (3) The numbers of vertices in different rectangular areas are relatively similar, which can effectively avoid the workload-imbalance issue. 
After the reordering, to fully utilize the resources of PL and AIEs, we will map the feature aggregation of the vertices in the dense rectangular areas and in the remaining areas onto the AIEs and the PL, respectively. Note that both computations can be performed completely in parallel.

\subsection{AIE-based Sparse Tensor Engine}
\label{subsec:sparse-tensor}
As introduced in Section \ref{subsec:gcn}, the computation mode of GCNs is two-phase matrix multiplication. The essence of matrix multiplication is multiply-accumulate (MAC) operations. 
Matrix multiplication can be further decomposed into vector operations. An AIE provides a floating-point 512 bits SIMD vector unit, particularly two intrinsic calls, \textsc{fpmac} and \textsc{fpmul}, for vector multiplication and accumulation operations on the vector unit. \textsc{fpmac} performs multiplication and accumulation for single-precision real number real times floating-point vectors. \textsc{fpmul} does multiplication for single precision real times real floating-point vectors. Those intrinsic calls are designed and optimized for dense matrix multiplication. 

After the graph reordering, the density of rectangular areas is still lower than 10\% based on our extensive profiling results. Thus, we propose a lightweight strategy that enables efficient SpMM on AIEs, which improves the computation efficiency by avoiding zeros be involved in the computation and fully utilizes the high-frequency, single-instruction-multiple-data AIEs.
It is worth noting that, without our work, SpMM on AIEs is much slower than running the corresponding dense GEMM directly.
Besides, we also use the row-wise SpMM and the traditional sparse storage format CSR to increase the generality of our sparse tensor engine. 

\begin{figure}[h]
    \centering
    \includegraphics[width=\linewidth]{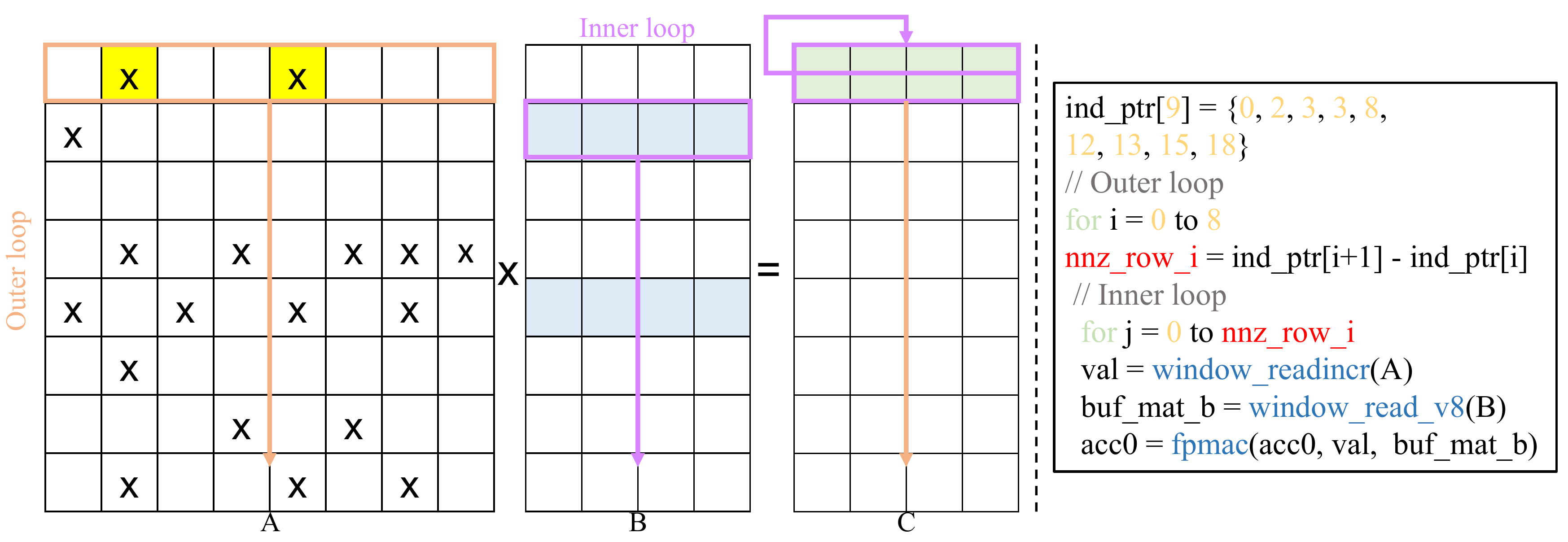}
    \vspace{-2mm}
    \caption{Row-wise sparse-dense matrix multiplication.}
    \label{fig:ori_spmm}
\end{figure}

Sparse row-wise product approach is all the non-zero elements from a single row of matrix $A$ are multiplied with corresponding rows of matrix $B$, where the row index of matrix $B$ is determined by the column index of the non-zero value from matrix $A$. The results are accumulated in the corresponding row of the output matrix (i.e., $C[i; :] = \sum_{k=0}^{N} A[i; k] \cdot B[k; :]$). Note that multiple rows can be computed in parallel.
Figure \ref{fig:ori_spmm} illustrates an example of row-wise SpMM.
The challenges of implementing row-wise SpMM include: (1) The number of the innermost loops is not fixed because the number of non-zeros in each row of matrix $A$ is not fixed. The compiler cannot use pipeline or loop flatten to optimize such loops with a variable number of loops, resulting in the final performance being worse than the dense matrix multiplication with the same size, even though we have theoretically reduced the amount of calculations. (2) CSR format leads to random row data accesses, which causes low memory bandwidth utilization. 

\begin{figure}[h]
    \centering
    \includegraphics[width=0.9\linewidth]{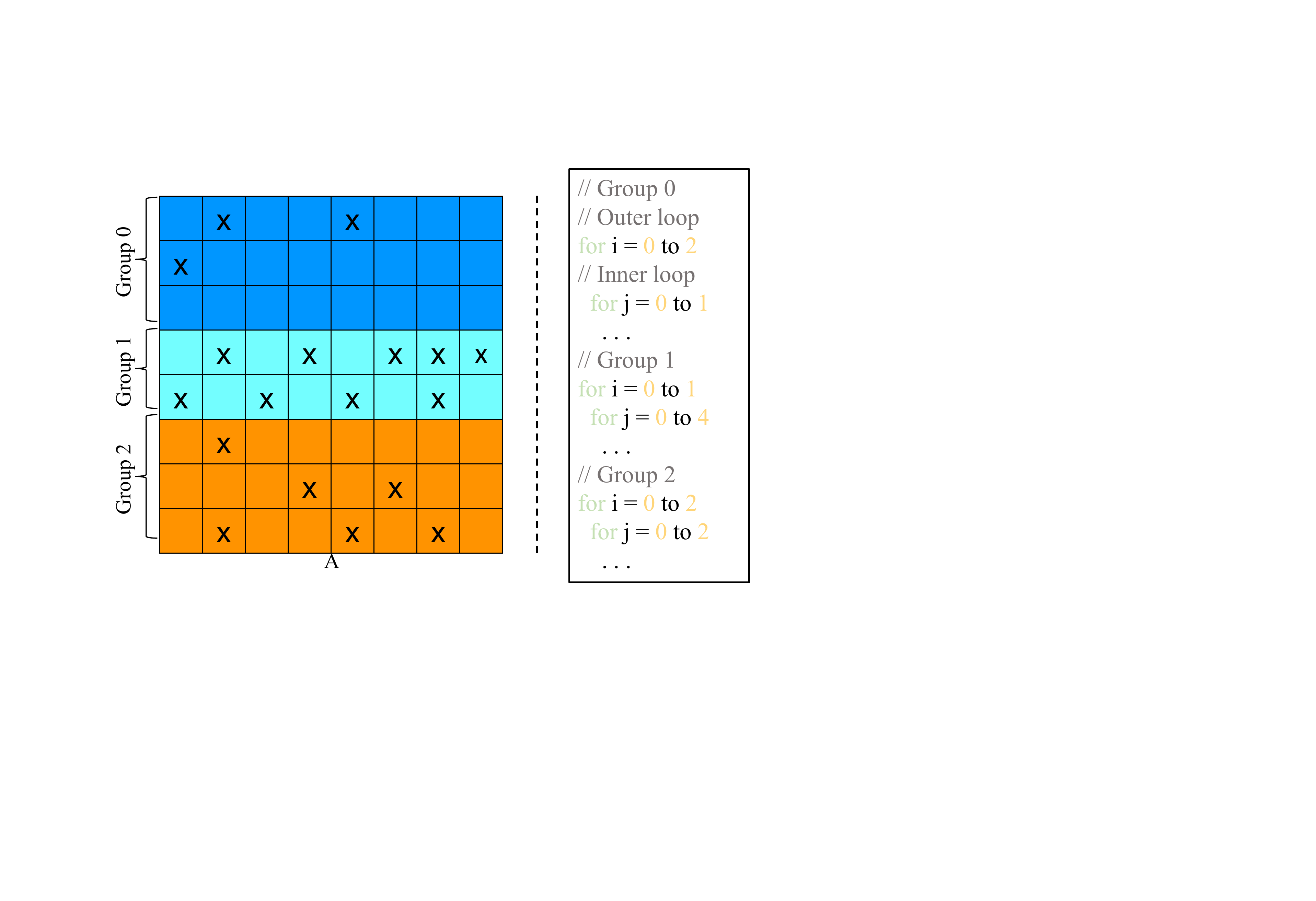}
    \vspace{-2mm}
    \caption{Grouped sparse-dense matrix and corresponding program.}
    \label{fig:new_spmm}
\end{figure}

We note that although directly flatten the outermost loop (each row of $A$ corresponding to a loop) can make the innermost loop fixed, each AIE has limited programming space, and direct expansion will cause compilation failures due to insufficient programming space. To solve this issue, we design a lightweight strategy (shown in Figure \ref{fig:new_spmm}) that divides the outermost loop into multiple loops with fixed number of innermost loops. This allows the compiler to fully optimize both loops. We propose to use ``moving average'' to divide the rows of matrix $A$ into multiple groups. Our goals include (1) each group contains as many rows as possible to save programming space, and (2) each group has as little padding as possible to reduce unnecessary calculations on zeros. 

\input{tex/algo1}

We describe our proposed grouping algorithm in Algorithm \ref{alg:grouping} in detail. We do not need to calculate the non-zeros of each line if $nnzs\_rows$ has already existed (lines 3-7). We use $pre\_ave$ to record the previous moving average, and $cur\_ave$ saves the current moving average (lines 8-9). 
Moreover, we also need to prevent dividing by zero since \textsc{reset()} function will set $cur\_ave$ to zero (line 15). If the change of the moving average exceeds threshold $\tau$, we put the data from row $j$ to row $i-1$ into a group, and we will pad each row in this group to ensure the same number of non-zero elements in each row, where $j$ is the first row of this group (lines 13-18). 

\subsection{Sparse Systolic Tensor Array on AIEs}
\label{subsec:sparse-systolic}
Two-dimensional (2D) systolic array is a pipelined 2D array of processing elements (PEs). 
Classical systolic array is generalized into a family of systolic tensor array by replacing the traditional scalar PEs with tensor PEs (TPEs). Each TPE is responsible for processing one tile of tensor or matrix. 
When using systolic tensor array to perform matrix operations, TPEs in the same row are required to perform exactly the same calculation mode (e.g. MAC) because one tile of data will flow to each TPE in the same row in turn. It is very easy to satisfy such requirements when performing dense matrix multiplication, because each TPE only needs to perform vector-based MAC operations.
But it is difficult to meet such requirements when performing SpMM using systolic tensor array, since each tile has a completely different number of non-zero element and computational model. 

To solve this issue, we propose an efficient method to process tiles of a sparse matrix to enable mapping SpMM onto the systolic tensor array automatically. Our idea is to pad the tiles in the same row as little as possible to make them have the same calculation pattern. Algorithm \ref{alg:TPE} describes the simplified workflow of automatic pre-processing of tiles and corresponding tensor PEs generation. We generate different sparse or dense codes for the systolic tensor PEs in the same row as the distributions of non-zeros in different tiles are different. 
Specifically, 
(1) we count the non-zeros of tiles in the same row (lines 6-8). 
(2) We calculate the average non-zeros ($ave\_nnz$) and maximum non-zeros ($max\_nnz$) of all tiles in the same row (lines 9-10). 
(3) We attempt to find a suitable number of non-zeros (line 12) for all tiles in the same row if the difference between $ave\_nnz$ and $max\_nnz$ is larger than the pre-defined ratio $\delta$; if we cannot find a suitable number, we will select $max\_nnz$ as ideal non-zeros for all tiles in the same row (line 14).
The purpose of this step is to reduce padding as much as possible. The function \textsc{find\_nnz} is to find the number of non-zeros which covers $p$ percentage of all tiles in the same row. The remaining non-zeros are calculated by SpMM in PL. 
(4) We use the grouping algorithm described in Algorithm \ref{alg:grouping} to group the rows (enable efficient SpMM on each AIE) after generating the number of non-zeros in each row (line 17), and obtain the final density after padding.
(5) We directly use dense tensor PE for those tiles if their final density is larger than $d$; otherwise, we use sparse tensor PE to process those tiles (lines 18-22). Based on our profiling experiments, there is no speedup of using spare tensor PE when density higher than 50\% (shown in Figure \ref{fig:csr}). 
\input{tex/algo2}

\subsection{Pipelining SpMM Chains}

\begin{figure}[!t]
    \centering
    \includegraphics[width=0.9\linewidth]{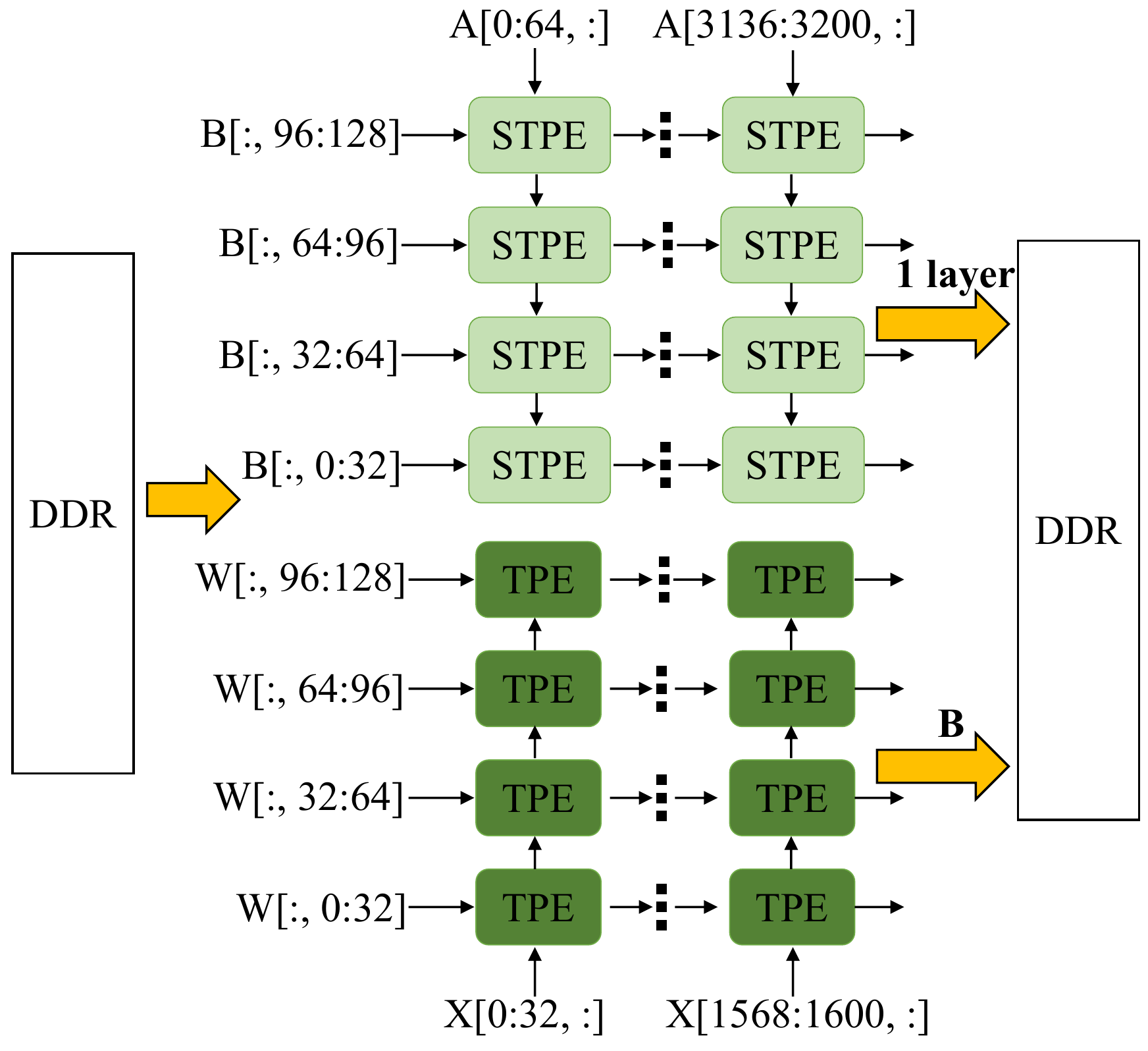}
    \caption{Our proposed computation mapping strategy and pipelining.}
    \label{fig:pipeline}
\end{figure}

\textbf{Intra-Layer SpMM Pipelining.} As described in Section \ref{subsec:gcn} and Equation \ref{equ-1}, SpMM chains $A \cdot (X \cdot W)$ are executed on three different hardware, i.e., dense systolic tensor array, sparse systolic tensor array, and PL for SpMM. Figure \ref{fig:pipeline} illustrates how to map such computation pattern onto the AIEs. Note that ``:'' means all indices along this axis. For instance, B[:, 0:32] means a slice from $B$ containing 32 columns across all the rows.
Note that there are 400 AIEs distributed in 8 rows and 50 columns. The upper 4 lines of the AIEs are used to implement the mixed sparse or dense systolic tensor PEs (STPEs/TPEs) to perform the computation of $A \cdot B$. We use Algorithm 2 to automatically generate corresponding STPEs/TPEs based on the sparsity. According to our experiment, over 90\% of generated systolic tensor PEs are sparse. The remaining 4 lines of the AIEs are used to implement the dense systolic tensor PEs to perform the computation of $X \cdot W$, where $B$ is the intermediate variable generated by $X \cdot W$.

The tile size (i.e., the size of a tile in the blocked matrix-matrix multiplication) of $A \cdot B$ (SpMM) is 64$\times$64.
The reasons for choosing this tile size are: (1) $A$ is represented in a CSR format, so such a tile of $A$ can be completely stored in the on-chip memory of an AIE. (2) Feeding a large amount of data can ensure the computation efficiency of the AIE. The tile size of $X \cdot W$ (dense matrix-matrix multiplication) is 32$\times$32, which is the maximum size that an AIE can hold after forming systolic tensor array. The remaining 4 lines of the AIEs will be reconfigured to STPEs/TPEs after finishing the entire $X \cdot W$, maximizing the use of all AIE resources. Note the matrix size equals the tile size multiplied by the number of tensor PEs.

Note that $A$ is constant during the inference of a certain graph, once a partial result $p_B$ of $B$ is calculated, we can start the multiplication of $p_B$ with $A$ on STPEs/TPEs and PL for SpMM immediately without waiting for the entire $X \cdot W$ to finish. Therefore, we can exploit the parallelism between consecutive SpMMs--$X \cdot W$ and $A \cdot (X \cdot W)$---in a layer through fine-grained pipelining, as shown in Figure \ref{fig:pipeline}. When generating a tile (i.e., 32$\times$32) of intermediate data $B$, we perform $A \cdot B$ immediately. This pipelining design has two major benefits: (1) It gains extra parallelism and reduce the overall latency. (2) It avoid a part of hardware stalls.


%% file: tex/algo1.tex
\setlength{\textfloatsep}{0pt}
\begin{algorithm}[!t]
\footnotesize\sffamily
\normalfont
\SetAlgoLined
\setcounter{AlgoLine}{0}
\SetKwComment{Comment}{\# }{}
\SetKwInOut{Input}{Inputs}
\SetKwInOut{Output}{Outputs}
\Input{$A$: input array; $nnzs\_rows$: non-zeros of each row; $rows$: the number of rows of $A$; $\tau$: threshold of changing group}
\Output{$group\_dic$: dictionary of group information; $density$: density after padding}
\BlankLine

$moving\_ave \gets$ MovingAverage(); $group\_dic \gets$ dict(); $idx\_g \gets 0$\par

\For{$i \gets 0$ \KwTo $rows-1$}{
    \eIf{not exist($nnzs\_rows$)}
    {
        $nnz\_row\_i \gets$ count\_nonzero($A[i, :]$)\par
    }
    {
        $nnz\_row\_i \gets nnzs\_rows[i]$\par
    }

    $pre\_ave \gets cur\_ave$\par
    $cur\_ave \gets moving\_ave.$update($nnz\_row\_i$)\par
    
    \If{$pre\_ave == 0$}{
      $pre\_ave \gets cur\_ave $ \hfill\Comment{\ttfamily Prevent division by 0.}
    }
    \eIf{$ abs(cur\_ave - pre\_ave) / pre\_ave \geq \tau $}
    {
        $group\_dic[idx\_g] \gets g$; $g \gets [\ ]$\hfill
        \Comment{update group.}
        $moving\_ave.$reset(); $moving\_ave.$update($nnz\_row\_i$)\par
    }
    {
        $g.$append($i$)
    }
}
 wi$density \gets \text{calc\_density}(group\_dic)$\par

\caption{Proposed grouping algorithm.}
\label{alg:grouping}
\end{algorithm}




    


%% file: tex/algo2.tex
\setlength{\textfloatsep}{0pt}
\begin{algorithm}[!t]
\footnotesize\sffamily
\normalfont
\SetAlgoLined
\setcounter{AlgoLine}{0}
\SetKwComment{Comment}{\# }{}
\SetKwInOut{Input}{Inputs}
\SetKwInOut{Output}{Outputs}
\Input{$A$: input sparse matrix; $rows$: the number of rows of $A$;  $cols$: the number of columns of $A$;  $tile\_size$: tile size; $\delta$: ratio by which the number of non-zeros changes. $p$: coverage percentage; $d$: density threshold of generating dense tensor PE.}
\Output{Sparse or dense code for systolic tensor PEs in the same row.}
\BlankLine

$tiles\_row = \frac{rows}{tile\_size}$; $tiles\_col = \frac{cols}{tile\_size}$\par
\For{$i \gets 0$ \KwTo $tiles\_row$}{
    $nnzs\_rows \gets [0] \times tile\_size$\par
    \For{$j \gets 0$ \KwTo $tile\_size$}{
        $nnzs\_row\_j \gets [0] \times tiles\_col$\par
        \For{$k \gets 0$ \KwTo $tiles\_col$}{
            $nnzs\_row\_j[k] \gets$ count\_nonzero($A[i \times tile\_size + j, k \times tile\_size : (k+1) \times tile\_size]$)\par
        }

        $ave\_nnz \gets \text{sum}(nnzs\_row\_j) / \text{len}(nnzs\_row\_j)$ \par 
        $max\_nnz \gets \text{max}(nnzs\_row\_j)$ \par

        \eIf{$\frac{max\_nnz}{ave\_nnz} \geq \delta$}
        {
            $nnzs\_rows[j] \gets$ find\_nnz($nnzs\_row, p$)
        }
        {
            $nnzs\_rows[j] \gets max\_nnz$
        }
    }
    
    $group\_dic, density \gets grouping(nnzs\_rows)$
    
    \eIf{$density \geq d$}
    {
        gen\_dense\_tensor\_PE($i$)
    }
    {
        gen\_sparse\_tensor\_PE($i, group\_dic$)
    }
}

\caption{Proposed automatic tensor PEs generation algorithm.}
\label{alg:TPE}
\end{algorithm}

%% file: tex/05_Evaluation.tex
\section{Experimental Evaluation}
\label{sec:evaluation}
In this section, we first introduce the experimental setup and analyze the performance impact of graph reordering and mapping methodologies. Then, we compare the performance of H-GCN with the state of the art GCN accelerators.

\begin{table}[b]
    \caption{Test graph datasets.}
    \centering
    \resizebox{0.8\linewidth}{!}{
    \input{tex/datasets}
    }
    \label{tab:datasets}
\end{table}

\begin{figure*}[h]
    \centering
    \includegraphics[width=\linewidth]{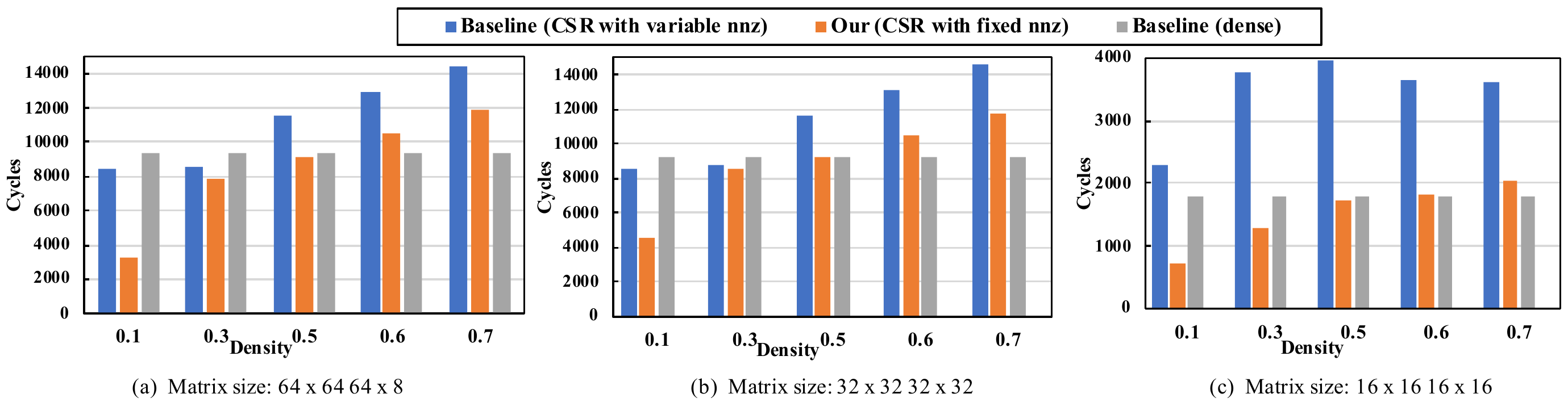}
    \caption{Speedups of sparse tensor engine with different grouping strategies under different matrix sizes.}
    \label{fig:csr}
\end{figure*}

\begin{table*}[h]
    \caption{Comparison of inference times (T) in $\mu$s and energy efficiency (E) in graphs/kJ. OoM is short for ``out of memory''.}
    \centering
    \resizebox{\linewidth}{!}{
    \input{tex/time1}
    }
    \vspace{-2mm}
    \label{tab:time1}
\end{table*}

\begin{table}[h]
    \caption{Comparison of inference times (T) in $\mu$s and energy efficiency (E) in graphs/kJ.}
    \centering
    \resizebox{\linewidth}{!}{
    \input{tex/time2}
    }
    \label{tab:time2}
\end{table}

\subsection{Experimental Setup}
\label{subsec:setup}

\textbf{Dataset.} Our Graph accelerator evaluation covers a widely used spectrum of mainstream graph datasets \cite{geng2021gcn, zhang2020hardware} including Cora \cite{yang2016revisiting}, Citeseer \cite{yang2016revisiting}, Pubmed \cite{yang2016revisiting}, Flickr \cite{zeng2019graphsaint}, Reddit \cite{zeng2019graphsaint}, Yelp \cite{zeng2019graphsaint}, and AmazonProducts (Amazon) \cite{zeng2019graphsaint}. Details of these datasets are listed in Table \ref{tab:datasets}. 

\textbf{GCN Model.} Similar to the previous works \cite{yan2020hygcn, geng2020awb}, we evaluate our solution on two-layer Vanilla-GCN model \cite{kipf2016semi} with the hidden dimension of 128.

\textbf{Our Platform.} We use Xilinx Versal VCK5000 (data center development card) \cite{vck5000} and its development kit for implementation. VCK5000 features the Xilinx Versal ACAP XCVC1902 device. XCVC1902 device contains 400 AIEs distributed in 8 rows and 50 columns. For PL resources, XCVC1902 device includes 1,968 DSP engines, 1,799,680 CLB Flip-Flops (FFs), 899,840 LUTs, and 34 MB Block RAM. VCK5000 board is equipped with four discrete DDR4 with 72-bit memory interface. The external memory has 100 GB/s peak memory bandwidth with four memory channels. Each channel can provide 25 GB/s peak memory bandwidth. 
We compile our design using Vitis unified software platform 2020.2.

\textbf{Baseline Platforms.} We compare our H-GCN with two advanced, well-optimized geometric deep learning frameworks, i.e., PyG \cite{fey2019fast} and DGL \cite{wang2019deep}, on general-purpose processors (i.e., CPU and GPU) and the state-of-the-art GCN accelerators, i.e., HyGCN \cite{yan2020hygcn}, AWB-GCN \cite{geng2020awb}, I-GCN \cite{geng2021gcn}, and BoostGCN \cite{zhang2021boostgcn}.
The CPU platform is equipped with two 28-core Intel Xeon Gold 6238R @2.2GHz processors with 384 GB DRAM. The GPU platform is equipped with an NVIDIA RTX 2060 SUPER with 8 GB memory. We denote PyG and DGL running on CPU and GPU platforms as PyG-CPU, DGL-CPU, PyG-GPU, and DGL-GPU, respectively. PyTorch version and CUDA version are 1.11.0 and 11.3, respectively. 

\textbf{Implementation Details.}
First, we map different partitioned computations to different engines as follows: (1) when the density is higher than 50\%, we map the computation of tightly clustered subgraphs onto dense AIEs; when the density is lower than 50\% but higher than 1.0\%, we map the computation of loosely clustered subgraphs onto sparse AIEs; and when the density is lower than 1.0\%, we map the computation of scattered nodes onto PL.
Second, we follow three steps to conduct this allocation: (1) we compile the code of AIEs for the computation of clustered or loosely clustered nodes (after reordering) using the Vitis AI compiler; (2) we compile the HLS kernels of PL for the computation of scattered nodes using the v++ command; and (3) we use the v++ command to link the compiled objects with the target platform (i.e., VCK5000).
Third, the frequency of NoC, PL, and AIEs is 800 MHz, 273 MHz, and 1GHz, respectively. The hardware resource utilization and frequency are obtained from the generated report by place-and-route. Note that the frequencies of PL and NoC are defined by our design choice, while AIEs–--an array of VLIW processors with SIMD vector units–--have a fixed frequency of 1 GHz.
Fourth, the SpMM module only accounts for 15.3\%, 84.6\%, 14.7\%, and 26.6\% of BRAM, DSP, FFs, and LUTs, respectively. 
Last, the evaluation results shown in the following discussion are based on simulations. Xilinx provides a profiling tool called ``Vitis Analyzer'' \cite{vitis_analyzer}, which can accurately model the execution time of AIEs.

\subsection{Speedup of Sparse Tensor Engine}
First, we evaluate the impact of the grouping algorithm on the overall speedup. We perform the experiments on different matrix sizes and densities as illustrated in Figure \ref{fig:csr}.
Since an AIE can only hold up to $64 \times 64 + 64 \times 8$ floating-point numbers, we test matrix sizes up to 64. 
Compared to the original dense algorithm, our grouping algorithm (i.e., CSR-fixed-nnz) provides 2.9$\times$, 2.1$\times$, and 2.5$\times$ speedup over the original dense method on matrices of size 64, 32, and 16, respectively, when density is 0.1. 

The row-wise SpMM with variable loops (i.e., CSR-variable-nnz), however, is much slower than the dense method even though we theoretically avoid computation on zeros. This is because the Vitis AIE compiler cannot use pipelining or loop flattening to optimize those variable loops.

The speedup gradually decreases to 1 as the density increases, and the speedup disappears when the density is higher than 50\%. The reasons are the increase in non-zero elements leads to increases in both the overhead of random access data and the computational delay. Thus, we switch to dense matrix-matrix multiplication when the density is higher than 50\%. 

We also evaluate the impact of sparsity on the effective FLOPS of an AIE. The effective FLOPS is 7.1 GFLOPS per AIE for dense matrix multiplication. We calculate the effective FLOPS based on nonzeros. FLOPS will increase as the density increases. This is because SpMM needs to convert to dense vector operations for executing on AIEs. For example, the effective FLOPS per AIE for SpMM of 32×32 by 32×32 is 1.6 GFLOPS, 2.5 GFLOPS, 3.1 GFLOPS, 3.4 GFLOPS, 3.5 GFLOPS, and 3.7 GFLOPS, when the density is 10\%, 20\%, 30\%, 40\%, 50\%, and 60\%, respectively.

\subsection{Comparison with State of The Art}
We evaluate the inference latency, and energy efficiency of H-GCN and compare it with other approaches (including software and accelerator solutions).

First, the ``T'' columns in Table \ref{tab:time1} show that H-GCN outperforms the best accelerator I-GCN by 1.1$\times$ in terms of inference latency. Moreover, compared with other prior accelerators, H-GCN provides speedups of 1.5$\times$$\sim$2.3$\times$ (1.9$\times$ on average) over BoostGCN, 1.2$\times$ over AWB-GCN, and 6.9$\times$ over HyGCN.
In addition, H-GCN significantly outperforms PyG and DGL on both CPU and GPU: it achieves average speedups of 79.5$\times$ over PyG-CPU, 12.2$\times$ over DGL-CPU, 1.59$\times$ over PyG-GPU, and 1.58$\times$ over DGL-GPU.

The performance improvement is because of (1) the better data locality and hence higher data reuse after the graph reordering, (2) the full use of AIEs via efficient sparse systolic tensor computation, and (3) our proposed scheduling approach for reducing the number of stalls in the overall pipeline.

The ``E'' columns in Table \ref{tab:time1} show that H-GCN is 1.12$\times$ and 1.64$\times$ more energy-efficient than I-GCN and AWB-GCN, respectively, which were previously the most energy-efficient solutions. This is due to the ACAP's more efficient dynamic power management \cite{vitis_power}. Note that we measure the energy efficiency of H-GCN by using Xilinx Power Estimator \cite{vitis_power}.

For relatively small graphs, dataflow accelerators such as I-GCN normally preload the graph data into their on-chip buffer and thereby avoid off-chip data access achieving lower inference latency. Therefore, we compare H-GCN with CPU and GPU platforms for Cora, Citeseer, and Pubmed. 
Table \ref{tab:time2} compares inference latency and energy efficiency of relatively small graphs in CPU and GPU platforms. It achieves average speedups of 71.1$\times$ over PyG-CPU, 59.8$\times$ over DGL-CPU, 10.9$\times$ over PyG-GPU, and 19.2$\times$ over DGL-GPU.

\subsection{Performance Breakdown}
To demonstrate that the performance improvement is due to the proposed method rather than the graph reordering, we map the computation of dense rectangular areas into AIEs without the approach (using dense systolic tensor array). The inference time of Cora, Citeseer, Pubmed, Flickr, Reddit, Yelp, and Amazon increases by 2.0$\times$, 2.9$\times$, 4.3$\times$, 5.9$\times$, 1.9$\times$, 4.3$\times$, and 3.9$\times$, respectively. 

We compare the performance of SpMM (i.e., 64$\times$64 by 64$\times$32) on PL and AIEs with different sparsities. Specifically, when the densities are 0.1\%, 0.5\%, 1.0\%, 5.0\%, and 10.0\%, the run times of PL are 0.18 $\mu$s, 0.88 $\mu$s, 1.75 $\mu$s, 8.41 $\mu$s, and 16.82 $\mu$s, respectively. The run times of AIE are 1.1 $\mu$s, 2.07 $\mu$s, 3.84 $\mu$s, 7.97 $\mu$s, and 10.44 $\mu$s, respectively.
This illustrates that SpMM on PL is faster than on AIE when the density is less than 1.0\%. Thus, we propose to use ``density'' as our criterion to determine whether to map SpMM onto PL or AIE.

In addition, we propose to prefetch and cache data through the PL controller because the theoretical PL-AIE bandwidth can reach 1.3 TB/s, whereas AIE-NoC bandwidth is only around 12 GB/s. Our evaluation shows that PL-DDR bandwidths of Cora, Citeseer, Pubmed, Flickr, Reddit, Yelp, and Amazon are 72.6 GB/s, 71.9 GB/s, 69.3 GB/s, 81.7 GB/s, 79.0 GB/s, 74.5 GB/s, and 75.7 GB/s, respectively.
Note that since Xilinx provides DDR controller IP, we implement our own DDR controller on PL. To calculate the throughput we use RTL simulations to measure the total clock cycles for transferring the graph data.

\subsection{Overhead of Graph Reordering}
\label{subsec:re-overhead}

Finally, we evaluate the time overhead of the graph reordering, as shown in Table \ref{tab:time3}. Note that as aforementioned, the graph reordering can be integrated into the training process \cite{chiang2019cluster}, so we take this overhead as the offline overhead. The OpenMP version of Metis takes advantage of multiple cores/threads in the CPU to reorder large graphs in parallel. For the Amazon dataset with 1,569,960 vertices, the graph reordering on 56 CPU cores only takes 7.31 seconds. 

\begin{table}[ht!]
    \caption{Graph reordering time ($m$s).}
    \centering
    \resizebox{\linewidth}{!}{
    \input{tex/time3}
    }
    \label{tab:time3}
\end{table}

Since graph can evolve dynamically, especially for inductive GNNs, we will support this online graph reordering in our future work. 
Specifically, we plan to use the host’s CPU to reorder the initial graph offline (by only once) and the ACAP’s ARM CPU to fine-tune the order online (by multiple times) as the graph evolves. This will help eliminate the communication cost of transferring node indices between the host and ACAP.

%% file: tex/datasets.tex
\begin{tabular}{@{} >{\bfseries}rrrr@{}}  
\toprule
    \textbf{\sffamily Dataset} 
    &\textbf{\sffamily \# Vertices}
    &\textbf{\sffamily A's Density}
    &\textbf{\sffamily \# Features} \\
\midrule
Cora     &2,708   &0.14\%   &1,433 \\
Flickr  &89,250     &0.011\%  &500\\
Citeseer &3,327   &0.08\%   &3,703 \\
Reddit  &232,965    &0.04\%   &602\\
Pubmed   &19,717  &0.023\%  &500 \\
Yelp    &716,847    &0.0027\% &300\\
Amazon  &1,569,960  &0.011\%  &200\\
\bottomrule
\end{tabular}

%% file: tex/time1.tex
\begin{tabular}{@{} >{\bfseries}rrr|rr|rr|rr|rr|rr|rr|rr|rr}
\toprule
    \multirow{2}{*}{\textbf{\sffamily Dataset}}
    &\multicolumn{2}{c|}{\textbf{\sffamily PyG-CPU}}
    &\multicolumn{2}{c|}{\textbf{\sffamily DGL-CPU}}
    &\multicolumn{2}{c|}{\textbf{\sffamily PyG-GPU}}
    &\multicolumn{2}{c|}{\textbf{\sffamily DGL-GPU}}
    &\multicolumn{2}{c|}{\textbf{\sffamily HyGCN}}
    &\multicolumn{2}{c|}{\textbf{\sffamily AWB-GCN}}
    &\multicolumn{2}{c|}{\textbf{\sffamily I-GCN}}
    &\multicolumn{2}{c|}{\textbf{\sffamily BoostGCN}}
    &\multicolumn{2}{c}{\textbf{\sffamily H-GCN (our work)}} \\
    &T &E &T &E &T &E &T &E &T &E &T &E &T &E &T &E &T &E \\
\midrule
\midrule

Flickr   &3.5E5 &17.37  &2.4E5 &25.43  &1.6E4 &3.51E2  &1.1E4 &5.1E2  &N/A    &N/A     &N/A   &N/A    &N/A   &N/A    &2.01E4 &N/A  &1.02E4 &1.0E3 \\
Reddit   &6.5E6 &0.83   &5.4E5 &11.26  &OoM   &N/A     &6.6E4 &87.07  &2.89E5 &5.17E2  &5.0E4 &1.5E2  &4.6E4 &2.2E2  &9.81E4 &N/A  &4.18E4 &2.46E2 \\
Yelp     &5.9E6 &1.03   &8.6E5 &7.09   &OoM   &N/A     &2.5E5 &23.12  &N/A    &N/A     &N/A   &N/A    &N/A   &N/A    &1.93E5 &N/A  &1.2E5  &85.85 \\
Amazon   &OoM   &N/A    &2.9E6 &2.1    &OoM   &N/A     &OoM   &N/A    &N/A    &N/A     &N/A   &N/A    &N/A   &N/A    &7.94E5 &N/A  &5.15E5 &19.93E \\
\bottomrule
\end{tabular}

%% file: tex/time2.tex
\begin{tabular}{@{} >{\bfseries}rrr|rr|rr}
\toprule
    \multirow{2}{*}{\textbf{\sffamily Method}}
    &\multicolumn{2}{c|}{\textbf{\sffamily Cora}}
    &\multicolumn{2}{c|}{\textbf{\sffamily Citeseer}}
    &\multicolumn{2}{c}{\textbf{\sffamily Pubmed}} \\
    &T &E &T &E &T &E \\
\midrule
\midrule
PyG-CPU &1.1E4 &5.36E2  &1.7E4 &3.65E2  &5.7E4  &1.07E2 \\ 
DGL-CPU &7.5E3 &8.08E2  &2.4E4 &2.50E2  &2.9E4  &2.07E2 \\ 
PyG-GPU &2.2E3 &2.55E3  &2.7E3 &2.16E3  &3.7E3  &1.53E3 \\ 
DGL-GPU &4.1E3 &1.39E3  &4.6E3 &1.23E3  &4.96E3 &1.15E3 \\
H-GCN   &1.1E2 &9.18E4  &2.9E2 &3.56E4  &1.03E3 &9.93E3 \\
\bottomrule
\end{tabular}

%% file: tex/time3.tex
\begin{tabular}{rrrrrrr}
\toprule
  \textbf{\sffamily Cora}
& \textbf{\sffamily Citeseer}  
& \textbf{\sffamily Pubmed}
& \textbf{\sffamily Flickr}
& \textbf{\sffamily Reddit}  
& \textbf{\sffamily Yelp}
& \textbf{\sffamily Amazon} \\
\midrule
11.5 &11.2 &33.6 &193 &648 &1650 &7310\\
\bottomrule
\end{tabular}

%% file: tex/06_Conclusion.tex
\section{Conclusion and Future Work}
\label{sec:conclusion}
The heterogeneity of graph structure is a significant factor in limiting the performance of GCN inference. Moreover, since typical graphs consist of tightly clustered subgraphs, loosely clustered subgraphs, and scattered nodes, it is not possible to use a unified hardware architecture/device to accelerate all parts of a GCN computation. To solve these issues, we propose H-GCN, an ultra-efficient, systolic tensor-based hardware accelerator, with heterogeneous computation paradigm to corresponding to GCNs. 
We leverage the heterogeneity of the Xilinx Versal ACAP to process those three types of subgraphs efficiently. Our broad experiments have demonstrated that, compared with a state-of-the-art FPGA accelerator, H-GCN achieves speedups of 1.1$\sim$2.3$\times$. 
In the future work, we will address computation of gradually evolving GCNs by exploiting online graph reordering by leveraging the ARM processors in the Versal ACAPs.